\documentclass[runningheads]{llncs}
\usepackage{graphicx}
\usepackage{algorithm}
\usepackage{algorithmic}
\usepackage{xcolor}
\usepackage{multirow}
\usepackage{amsmath}
\usepackage{booktabs}
\usepackage{adjustbox}
\usepackage{textcomp, gensymb}
\usepackage[numbers]{natbib}
\usepackage{appendix}

\begin{document}

\title{
Clustering Running Titles to Understand the Printing of Early Modern Books} %
\titlerunning{Clustering Running Titles in Early Modern Books}
\author{Nikolai Vogler\inst{1}
\and
Kartik Goyal\inst{2}
\and
Samuel V. Lemley\inst{3}
\and
D.J. Schuldt\inst{4}
\and
Christopher N. Warren\inst{3}
\and
Max G'Sell
\and
Taylor Berg-Kirkpatrick\inst{1}
}

\authorrunning{N. Vogler et al.}
\institute{University of California San Diego \and
Georgia Institute of Technology \and Carnegie Mellon Univesity \and Simmons University \\
\email{\{nvogler,tberg\}@ucsd.edu, kartikgo@gatech.edu, samlemley@cmu.edu, d.j.schuldt@gmail.com, cnwarren@andrew.cmu.edu}}
\maketitle              %
\begin{abstract}
We propose a novel computational approach to automatically analyze the physical process behind printing of early modern letterpress books via clustering the running titles found at the top of their pages. 
Specifically, we design and compare custom neural and feature-based kernels for computing pairwise visual similarity of a scanned document's \textit{running titles} and cluster the titles in order to track any deviations from the expected pattern of a book's printing. 
Unlike body text which must be reset for every page, the running titles are one of the static type elements in a \textit{skeleton forme} i.e. the frame used to print each side of a sheet of paper, and were often re-used during a book's printing. To evaluate the effectiveness of our approach, we manually annotate the running title clusters on about 1600 pages across 8 early modern books of varying size and formats.
Our method can detect potential deviation from the expected patterns of such skeleton formes, which helps bibliographers understand the phenomena associated with a text's transmission, such as censorship. We also validate our results against a manual bibliographic analysis of a counterfeit early edition of Thomas Hobbes’ \textit{Leviathan} (1651) \cite{warrenetal2021}.\footnote{Code and data available at \url{https://github.com/nvog/clustering-running-titles}.}
 
\end{abstract}

\section{Introduction}

\begin{figure}[t]
    \centering
    \includegraphics[width=\textwidth]{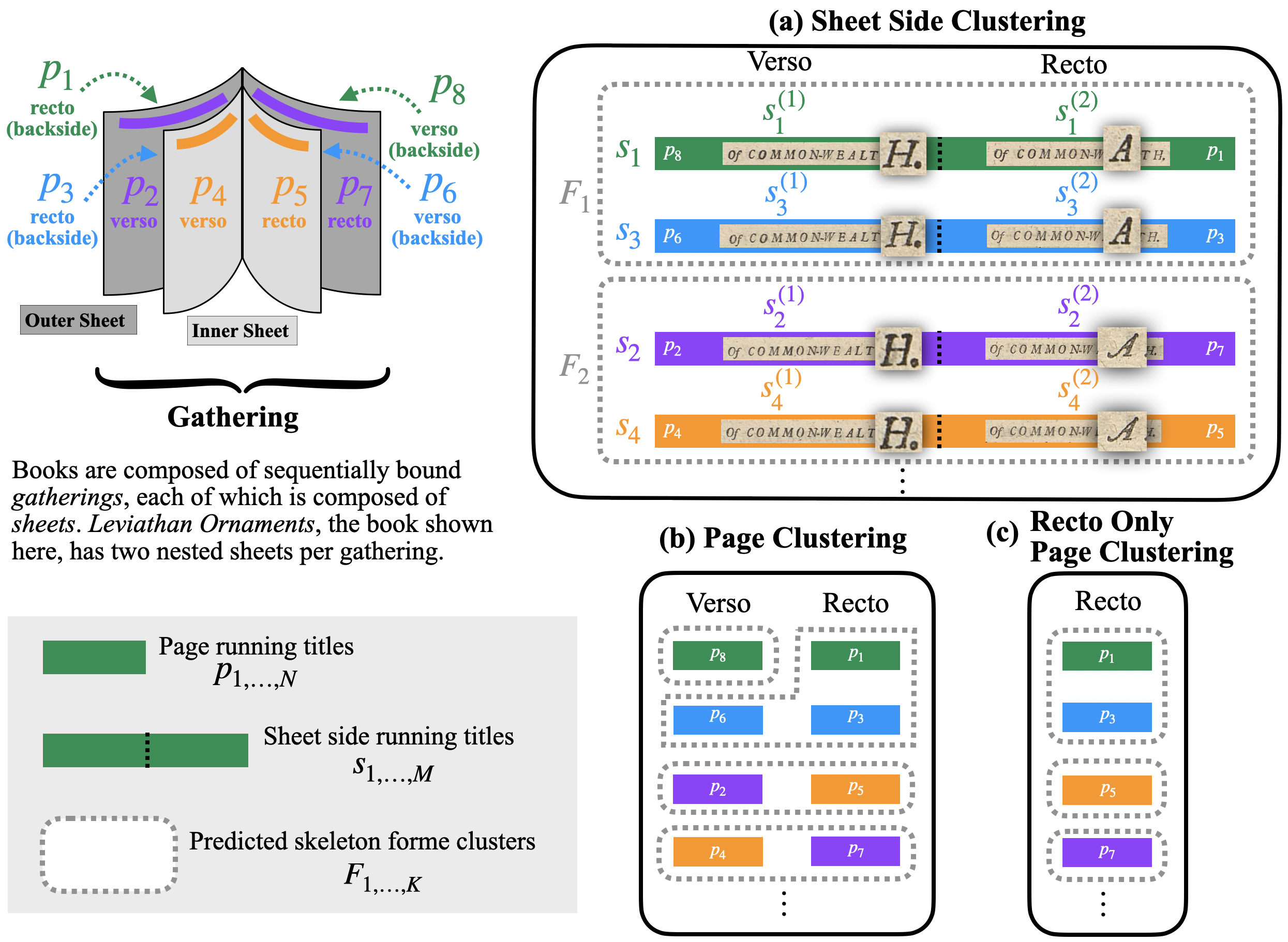}
    \caption{In this paper, we propose a new task that clusters running titles (colored rectangles) of early modern books into the underlying skeleton formes (outlined in gray dotted line) used to print them. 
    We show that by leveraging additional information about how the book was printed, namely the gathering structure of the book, we can cluster \textit{sheet sides} instead of \textit{pages} or \textit{recto pages} (i.e., right side pages), which greatly improves performance.}
    \label{fig:clustering_overview}
    \vspace{-2em}
\end{figure}

\begin{figure}[t]
    \centering
    \includegraphics[width=0.95\textwidth]{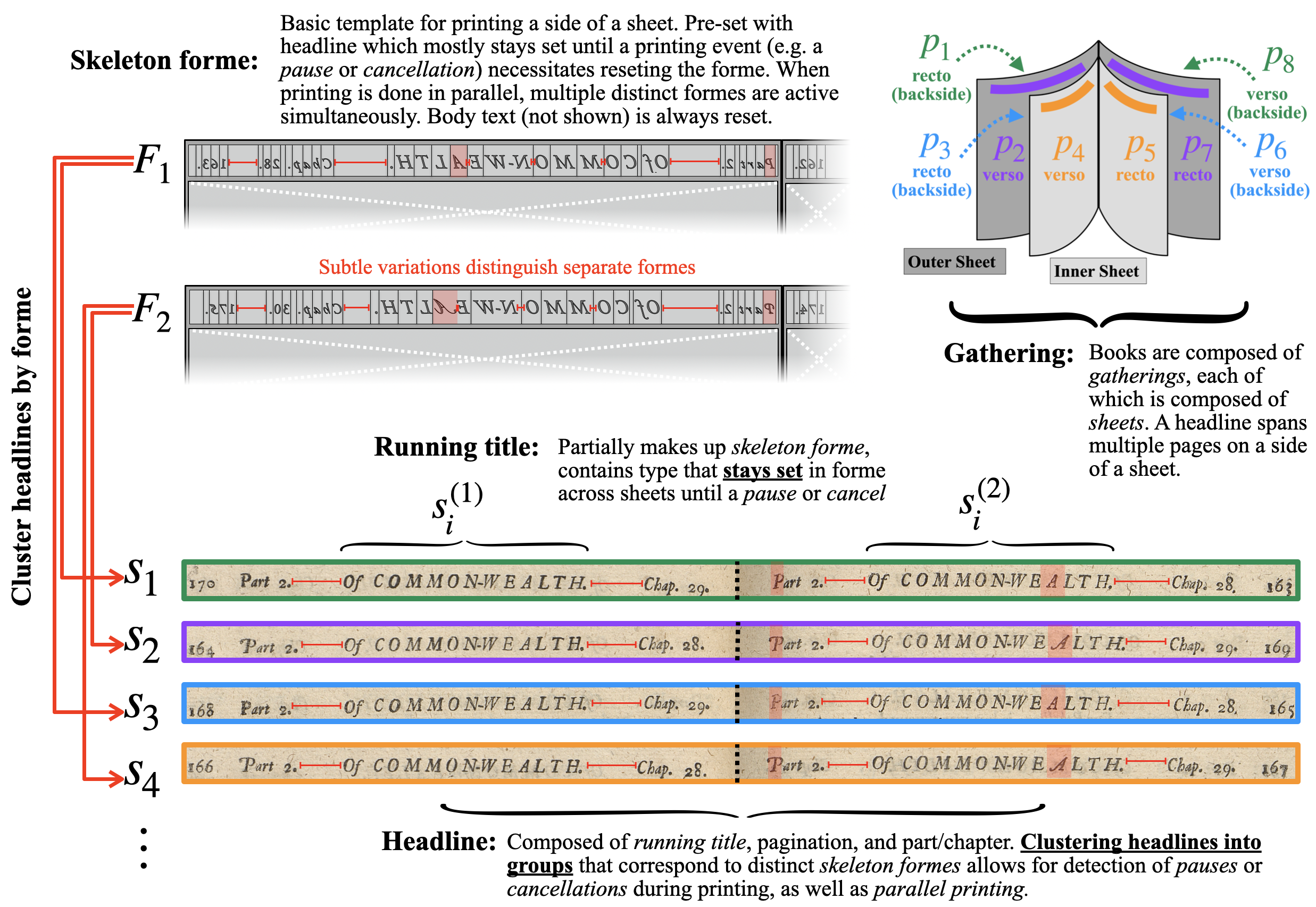}
    \caption{
    ``Skeleton forme'' and function in early modern printing. 
    On the left, two schematic metal skeleton formes $F_1$ and $F_2$, used in the printing of \textit{Leviathan}, are shown. In red, we highlight subtle differences in glyph anatomy and spacing between the headlines of $F_1$ and $F_2$. Underneath this, we display real groups of headlines from \textit{Leviathan} that are printed using the schematic skeleton formes $F_1$ and $F_2$ found by clustering on such spacing. 
    Each group is printed from the same skeleton forme with consistent running titles. 
    At the top right, we show how this book was folded and nested into constituent gatherings, with color-coded headlines corresponding to the actual headlines underneath.
    We note that body text is not shown because it is reset while printing.}
    \label{fig:overview}
    \vspace{-2em}
\end{figure}

Increasingly, AI methods are being applied to historical texts in order to enrich our knowledge about past cultures and societies, ranging from ancient Greek inscriptions \citep{assael2022restoring} to Akkadian tablets \citep{lazar2021filling} to Latin \citep{bamman2020latin}, early modern English and Islamicate books and manuscripts \citep{Vogler2022Lacuna}, inter alia \citep{sommerschield2023machine}.
Recent approaches \citep{ryskina2017automatic,Vogler2023Contrastive} focus on analyzing the printing process for early modern English (c. 1500--1800) books, which were produced on movable type printing presses and contain visible traces of their manual and material origins. 
One major source of this material evidence is found in so-called \textit{skeleton formes}, which are wooden frames containing bordering text and spacing material that was reused or even left in the printing press during a book's printing to ensure that the dimensions and layout of a book's pages remained constant. 
Accordingly, most of the typographic content in the bordering frame, such as the running title portion of the headline occurring at the top of a printed page, is expected to recur in subsequently printed sheets of the book. 
Type set in the skeleton forme lies in contrast to the body text, which was continuously reset for each new page.
Analysis of subtle changes occurring in the skeleton formes throughout a book can reveal how the book was printed, such as how many printing presses were involved and any potential censorship that may have occurred to the book \cite{bowers1949principles}.

In this paper, we describe an automatic, unsupervised computational approach to visual analysis of running titles in early modern books.  Our approach automatically clusters headlines extracted from images of the books of interest into groups that were printed using the same forme to amass evidence of printing phenomena. 
Distinctions in headlines that indicate variation among forme clusters are subtle, and specifying a robust feature representation for clustering is challenging. 
Hence, we design domain-informed specialized similarity kernels for clustering that are sensitive to the relevant variations and robust to other sources of noise. Specifically, we propose two similarity kernels, one based on Levenshtein edit-distance with a discretized representation of running title image intensities, and the other parametrized by a neural Vision Transformer cross-encoder fine-tuned on carefully generated synthetic data. These similarity kernels are then used to perform spectral clustering that yields the output for analysis.
Further, we leverage knowledge of each book's segmentation into gatherings and sheets in order to aggregate similarity scores across individual running titles from pages that fall within the same sheet, increasing the robustness of our method to both subtle variations and confounding noise.  
We quantitatively evaluate the predictions against a manually annotated suite of eight separate clustering problems comprising about 1600 running titles and found that both of our approaches outperform random baselines. Specifically, our simple domain-specific edit-distance based approach is very effective at this task and significantly outperforms the strong Vision Transformer-based cross-encoder approach. 
Also, we qualitatively analyze our results against a recent manual study of the printing of Thomas Hobbes' \textit{Leviathan} \cite{warrenetal2021}.

\section{Background: The Making of Early Modern Books}\label{sec:earlymodernprinting}

The Gutenberg-era printing press was set up to print onto large sheets of paper that were folded to size and bound together in sequence to make a book. 
The \textit{format} of the book was determined by the number of pages printed on each sheet of paper. 
The top right of Figure~\ref{fig:overview} shows a schematic of a single gathering in \textit{folio}---the simplest (though most expensive) format employed by early modern printers, as it required only a single fold and printing only two pages per side of a sheet.\footnote{In contrast, \textit{quarto} format involved printing four pages per sheet side, \textit{octavo} format eight pages per sheet side, etc. 
See Appendix Fig.~\ref{fig:quarto_octavo} for skeleton forme diagrams of these formats. Each format, diminishing by size, required an increasingly complex pattern of \textit{imposition}, or the arrangement in which sheets were set on the press to yield a sequentially-ordered gathering of pages after folding.}
Each printed page of a folio is located on either the \textit{recto} (i.e., front/right-side) or \textit{verso} (i.e., back/left-side) of a \textit{leaf}, which represents half of a full sheet.
Due to the common nesting arrangement of the gathering's two sheets, the first leaf of the gathering (from sheet 1, pages 1 \& 2) actually appears contiguously with the second leaf of the gathering (from sheet 2, pages 3 \& 4). 
This means that each gathering was printed in non-sequential page-pairs---namely, pages 1 and 8, 2 and 7, and 3 and 6, with 4 and 5 at the centerfold being the sole sequential exception. 
When set up on the press, each of these page-pairs made up a single forme, framed by a recurring skeleton.\footnote{For a detailed introduction to early printed books, refer to \citep{werner2019studying}.} 

In Figure~\ref{fig:overview}, we illustrate the early modern folio printing process with particular focus on how typesetting distinctions in the skeleton formes surface in the book after printed sheets are folded and gathered together.
First, compositors set individual pieces of type into formes (top left), which are then inked and relief printed onto paper, as shown in actual printed sheets from Thomas Hobbes' \textit{Leviathan} below formes $F_1$ and $F_2$.
Next, sheets must be folded and formulaically nested together into gatherings to assemble sequential components of the book (top right).
Finally, gatherings are bound together in order to complete the book. The two formes in this example would lead to two distinct clusters of running section titles containing the same text \textsc{Of Commonwealth}, but primarily differentiated by the character shapes and the spacing between the characters in the running section title. Hence, finding such clusters of running titles based on subtle visual differences is considered reliable evidence (top left, in red) to conclude that multiple formes of the same running titles were involved in printing. Knowing, for example, that multiple formes were used simultaneously provides evidence about the size of a print shop, if sheets are interleaved, or of possible distributed printing when they are separated.
Further, in many cases knowing when a forme gets reset can suggest a pause in printing and even a cancellation (a reprinting of a previously-printed page) which can indicate the intervention of a censor or editor at the time of printing. Thus, the relevance of this narrowly focused kind of bibliographic analysis is surprisingly broad and can potentially bring new light to bare on the history of language and censorship.

The forme---not the page---was the basic component of early modern presswork; and the page order of early printed books was established by folding sheets. 
This matters for our analysis because while patterns in headlines recur at regular intervals throughout a book, facing pairs of headlines are separated, in most cases, by several intervening pages. %
Figure 2 shows that the pair of headlines in the outermost forme (pages 1 and 8) of the gathering is separated by six pages, for instance. 
Manually examining the skeleton forme shared by any two given pages is therefore involved, unintuitive, and impractical at scale. However, as we show in experiments, encoding knowledge of which pages fall on the same side of sheet into our computational model allows for more accurate clustering of pages into groups set by the same forme -- our primary analytical goal.

\begin{figure}[t]
    \centering
    \includegraphics[width=1.0\textwidth]{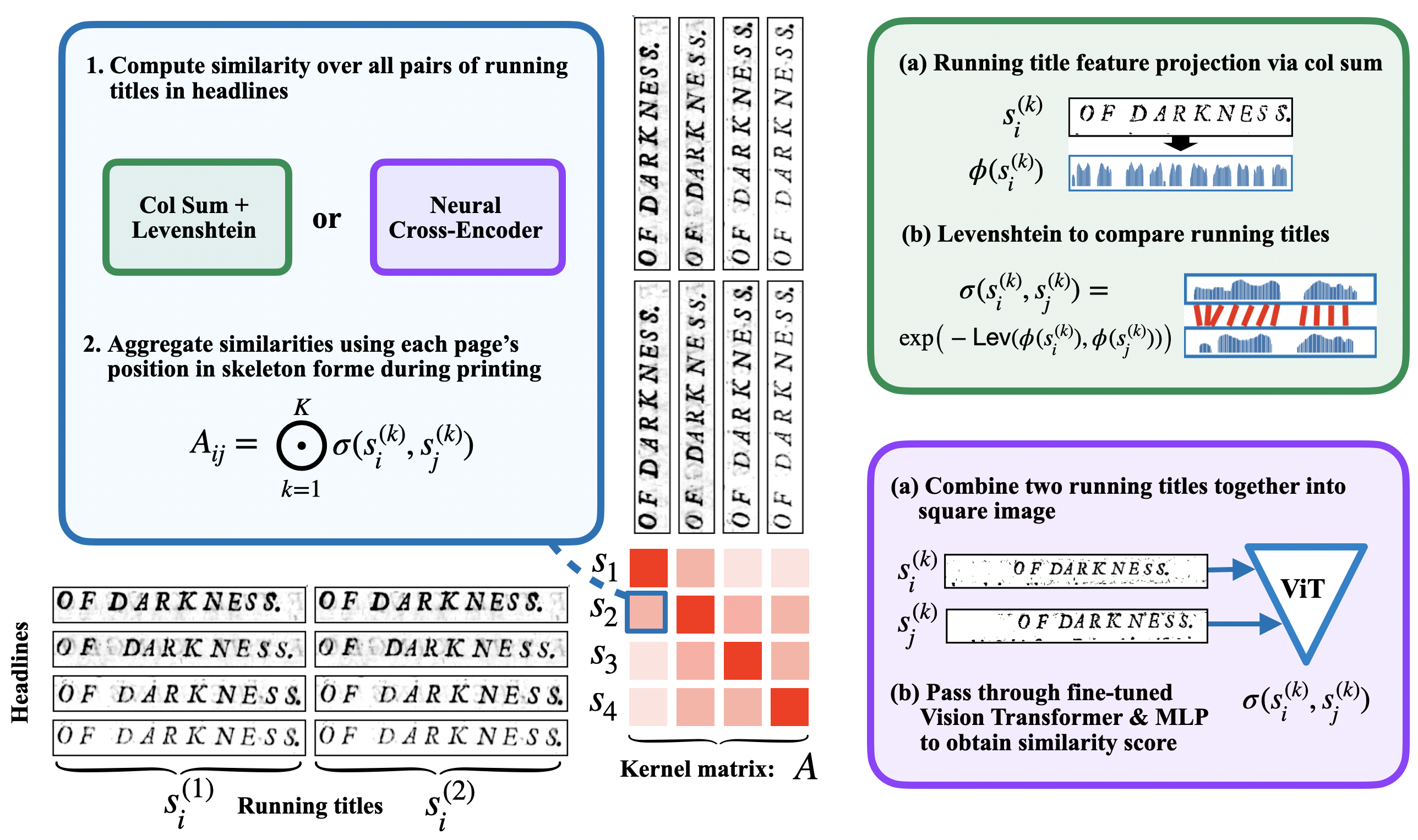}
    \caption{
    Proposed custom feature-based Levenshtein and neural cross-encoder Vision Transformer-based similarity kernels for comparing running titles of sheet sides in early modern printed books.
    We compute pairwise similarities between each \textit{corresponding running title} image on different sheet sides across the entire book, where $s_i$ and $s_j$ denote the two sheet sides being compared, with each of them containing $n$ running titles whose position and number is predetermined by the format of the book. 
    Here, $s_i^{(k)}$ represents the $k$-th positional running title on sheet $s_i$. 
    The kernel functions combine $k$ similarity computations performed between the running titles at the corresponding positions across sheet sides $s_i$ and $s_j$ via reduction operation $\bigodot$ (see Sec.~\ref{sec:kernels}).
    Using the similarities, we cluster kernel matrix $A$ to discover sheet sides printed with the same underlying skeleton forme. Running titles from `Of Darkness' section of \textit{Leviathan} are shown binarized.
    }
    \label{fig:methods}
    \vspace{-2em}
\end{figure}

\begin{table}[t]
    \centering
        \caption{The eight clustering problems in our proposed dataset. The first three columns denote number of datapoints for each clustering setting (all pages, recto, and sheet sides) and the \emph{fourth column represents number of ground truth clusters}. Experts determine this figure by manually comparing headline images from each book, as shown in Figure~\ref{fig:king_lear_clustering}. The `Format' column refers to the number of pages printed per side of a sheet (for instance, \textit{Leviathan} was printed 2 pages to a sheet) and number of leaves in the gathering (\textit{Leviathan} has 4, for a total of 8 pages per gathering).  For more on this, see Fig~\ref{fig:overview}.}
    \label{tab:dataset}
    { \setlength{\tabcolsep}{3pt} \begin{adjustbox}{width=\linewidth,center}
    \begin{tabular}{l | r r r r r r}
    \toprule
       Book  & \# Pgs. & \# Recto & \# Sheet  & \# \textbf{Skele. forme} &  Format & \# Leaves / \\
        &  & Pgs. & Sides & \textbf{clusters} &   & gathering \\
    \midrule
    \textit{Leviathan} & 376 & 188 & 94 & 20 & folio & 4\\
    \textit{Paradise Lost} & 336 & 168 & 84 & 12 & quarto & 4 \\
    \textit{King Lear} & 48 & 24 & 12 & 4 & quarto & 4  \\
    \textit{Mayor} & 72 & 36 & 18 & 6 & quarto & 4 \\
    \textit{Parthenissa} & 248 & 124 & 62 & 19 & quarto & 4\\
    \textit{Institution} & 80 & 40 & 10 & 2 & octavo & 8 \\
    \textit{Discourse} & 192 & 96 & 24  & 2 & octavo & 8\\
    \textit{Wisdom} & 240 & 120 & 30 & 8 & octavo & 8\\
    \bottomrule
    \end{tabular}
    \end{adjustbox}
    }
\end{table}

\section{Approach: Spectral Clustering with Custom Kernels for Running Title Variation}
As described above, since the skeleton formes are typically reused for unchanging running titles across several sheet sides, any deviation in the running titles across such sheet sides could indicate abnormal printing behavior, such as stop press, or cancellations due to errors or censorship. 
Therefore, \emph{we choose a sheet side to be our unit of quantitative analysis}. 
For our analysis, we pose the problem of identifying resetting of the formes as a clustering problem over all comparable sheet sides.
However, building a useful featurized representation of sheet sides, which is a prerequisite for common clustering algorithms like $k$-means, is difficult because the variation we are interested in characterizing includes very subtle differences between individual character shapes and spacing between them. We approach this problem by learning \emph{custom-designed kernel functions} that characterize variations in the running titles across sheet sides that yield similarity estimates between sheet sides that are useful for our clustering objective.
We describe two major approaches to estimate these suitable kernel functions below that characterize spacing and the shape differences we are interested in capturing.
Then, spectral clustering \cite{von2007spectral} is a natural approach to directly use this similarity matrix for clustering groups of sheet sides based on differences described by our kernel.

\subsection{Kernels for Analyzing Kerning  Variation}\label{sec:kernels}

In this section, we describe our approaches to establish a measure of visual similarity between sheet sides---our unit of analysis for running title clustering. 
Since the running titles are the only part of the page that remains constant across unchanging formes, we represent a sheet side by a collection of running titles belonging to the sheet (see Fig~\ref{fig:clustering_overview}). 
In order to obtain this representation, we preprocess page images to extract the running titles, and then use the information about book format to form groups of pages/running titles that belong to the same sheet side.
Formally, a sheet side $s$ contains multiple positional running titles $t$ depending on the format---for example, a folio sheet side consists of two running titles, one on the recto position and another on verso, and a quarto sheet side contains 4 running titles in a pre-specified positional order, and so on for other formats. 
As shown in Figure~\ref{fig:methods}, for comparing sheet sides containing $n$ positional running titles, we define a general kernel function $\kappa(s_i, s_j; \phi, \sigma)$ over sheet sides $s$, capturing similarity between them as: 
\begin{equation}
    \kappa\left(s_i, s_j; \phi, \sigma\right) = \bigodot_{k=1}^K\sigma\left(\phi(s_i^{(k)}), \phi(s_j^{(k)}) \right).
\end{equation}
As an example, Figure~\ref{fig:methods} depicts the computation of a similarity kernel over $4$ sheet sides of a folio, each of which have two running titles -- one verso and another recto.
In general, $s_i$ and $s_j$ denote the two sheet sides being compared, with each of them containing $n$ running titles whose position and number is predetermined by the format of the book. $s_i^{(k)}$ represents the $k$-th positional running title on sheet $s_i$. 
This kernel function essentially combines $k$ similarity computations performed between the running titles at the corresponding positions across the sheet sides $s_i$ and $s_j$ via a reduction operation $\bigodot$. The key motivation behind this formulation is that \emph{given knowledge of the format of a book and how it was gathered together, running titles across different positions on a sheet side do not affect each other because they must be printed by different parts of the skeleton forme}. For example, in Figure~\ref{fig:methods}, we use the knowledge about the folio format to identify verso and recto pages so that we never erroneously compute similarity between the verso and recto running titles. However, we still need a reduction operator that combines the similarity information across all the individual title positions on the sheet sides under comparison.
The individual similarity computation between the positional running titles is performed with function $\sigma$ applied on a pair of 
running title images $s_i^{(k)}$ and $s_j^{(k)}$ transformed individually by the feature map $\phi$. Below, we propose two parametrizations of the $\sigma$ and $\phi$ functions for the task of estimating similarity between two running titles: a custom-designed quantized Levenshtein kernel estimator (Lev) and a neural cross-encoder Vision Transformer-based (ViT) similarity function learned via domain-informed data augmentation.   

\paragraph{\textbf{Quantized Levenshtein Kernel (Lev):}}
While $\phi$ could simply be an identity map, which yields pixel-wise comparison between raw images, manual inspection of \textit{Leviathan}'s running titles suggests that a feature map that considers the kerning, or spacing between letterforms, would be a more distinguishing feature of running titles printed with the same skeleton forme. 
After experimentation with various feature functions $\phi$ and similarity function $\sigma$ combinations, we made a choice of the following featurization and similarity functions, which we collectively refer to as \emph{Quantized Levenshtein Kernel}. 
For the feature function $\phi$, as shown in Fig.~\ref{fig:methods}, we transform an image into a 1D column vector by summing over its height dimension, and then discretizing the columns into a set number of bins, where bin edges are determined either uniformly, by k-means, or via quantiles.
For the similarity function $\sigma$, we use the negative Levenshtein edit distance \cite{wagner1974string} between the two discretized 1D input strings. 
The Levenshtein distance between the string representations for two titles approximately capture the spacing differences and differences in character shapes in an automatic and robust manner because the column sums essentially represent the inking intensity across the width of the running title. 
Hence, this combination provides a good estimate of variation in kerning and also handles small amounts of noise encountered during the automatic extraction of running titles.

\paragraph{\textbf{Neural Cross-Encoder Vision Transformer Kernel (ViT):}} While the custom-designed edit distance-based kernel is aimed at capturing the subtle variations in spacing between character types and font shapes, the associated discretization step could cause useful features to be lost. 
Hence, we also devise a scheme to learn a flexible similarity metric between the running titles via a neural parametrization. 
Specifically, we fine-tune a pretrained Vision Transformer model \citep{dosovitskiy2020image}\footnote{We use the torchvision implementation with pretrained ImageNet weights and patch size of 16.} using either the triplet margin loss \cite{weinberger2009distance} or a binary log loss with negative sampling objective \cite{mikolov2013distributed} inspired by noise contrastive estimation \cite{gutmann2010noise} on carefully generated synthetic data
\begin{equation}
\min_\theta -\log p_\theta(s^{*} = s^{+}) - \frac{1}{k}\sum_{s^{-} \in \mathcal{S}^{-}} \log \left(1-p_\theta(s^{*} = s^{-})\right).    
\end{equation}
Here $s^{*}$ and $s^{+}$ are the running title images in the training data that match and should belong to the same forme. $\mathcal{S}^{-}$ denotes a set of $k$ similar-looking but non-matching running titles in the training data. 
$\theta$ denotes the parameters of the vision transformer that is being fine-tuned. 
This loss function maximizes the probability of correctly predicting whether the input running titles are matching or non-matching.

We use synthetic data for fine-tuning because the training objective requires supervision in the form of annotated matching pairs of running titles. 
Since we don't have such data for supervision, we generate training data by rendering the running title text via Early Modern print-inspired font libraries\footnote{Obtained from \url{www.onlinewebfonts.com}.} and perturbing the resulting images with noise to form positive and negative examples. 
Specifically, we want the neural similarity kernel to be robust to ineffectual perturbation like global offsets, inking, and other spurious noise sources. 
Hence, we create a \emph{matching} (positive) pair of titles by rendering an anchor image and perturbing that anchor image with such spurious noise as globally shifting the whole text randomly by up to 15\% of the width, skew, etc. 
We also want the neural kernel to be sensitive to subtle spacing variations between the character types of the running titles, and character shape variations among titles. 
Hence, for a rendered anchor image, we also generate similar looking but \emph{non-matching} (negative) titles by either adding spacing randomly between rendered characters in the anchor (instead of a global offset described above for the matching pairs), or also randomly swapping the font shape of up to 10\% of the letters with an italic/swash version in the same font family. 
Instead of changing the shape of a character, we also rarely replace a character with another similar looking character. 
For example, a period can be randomly replaced by a comma. The upshot of synthesizing the positive matching and negative non-matching pairs is that we can encourage the learning of a similarity function that is sensitive to the subtle shape and spacing variations we care about while remaining robust to other significant sources of variation like global offsets, inking variations etc. that do not indicate dissimilarity in the formes.

Additionally, because the pretrained vision transformer takes $224 \times 224$ square images as input, instead of padding the wide rectangular titles and learning an embedding for each item in the pair separately, we propose a scheme for a fused square representation of the input pair of titles which we call \emph{hack 'n stack}. 
As shown in Figure~\ref{fig:methods} (bottom right), we chop both wide running titles into $5$ parts and stack them vertically. 
Then, we concatenate these stacks to obtain a $224 \times 224$ square image which is passed as the input to the neural vision transformer. 
The transformer produces a similarity score $\sigma$ for this input image pair that is learned by the training objective above. 
Hence, we are able to learn an effective neural similarity kernel for our specialized purpose in a completely unsupervised manner via a domain-informed data augmentation scheme.

\paragraph{\textbf{Sheet side similarity via reduction:}} Finally, given similarity computation of corresponding running titles across all the positions on a sheet side with the combination of $\phi$ and $\sigma$ operators described above, we combine these measures via a reduction function $\bigodot$. 
For this reduction function, we experimented with norm functions with a range of orders $p$ $\left( ||\cdot||_p = \sqrt[p]{\sum_{k=1}^n |-\sigma(\phi(x^{(k)})|^p} \right)$. 
A higher value of $p$ has the effect of characterizing the distance between two sheet sides primarily by the position in which the corresponding running titles differ the most ($||\cdot||_\infty = \max(\cdot)$), while a lower $p$ focuses on the distances between the running titles at all positions more equitably.

\subsection{Spectral Clustering of Sheet Sides}

With similarity kernels that measure the visual relatedness of sheet sides as described above, we can construct square affinity matrices for all  comparable sheet sides in a book.
As shown with our sheet side setup in Fig~\ref{fig:methods}, each cell in this affinity matrix contains the aggregated similarity scores between a row/column sheet side pair, which can be viewed as edge weights connecting the sheet sides in a graph representation of the book.
We choose to use spectral clustering, a similarity-based clustering method, to turn this affinity matrix into a sparse binary adjacency matrix via thresholding, yielding the Laplacian matrix \cite{von2007spectral}.
Spectral clustering then performs $k$-means on the low-rank spectral decomposition of this  normalized Laplacian matrix.

\section{Dataset: A Suite of Skeleton Forme Clustering Tasks}\label{sec:dataset}

\begin{figure}[t]
    \centering
    \includegraphics[width=\textwidth]{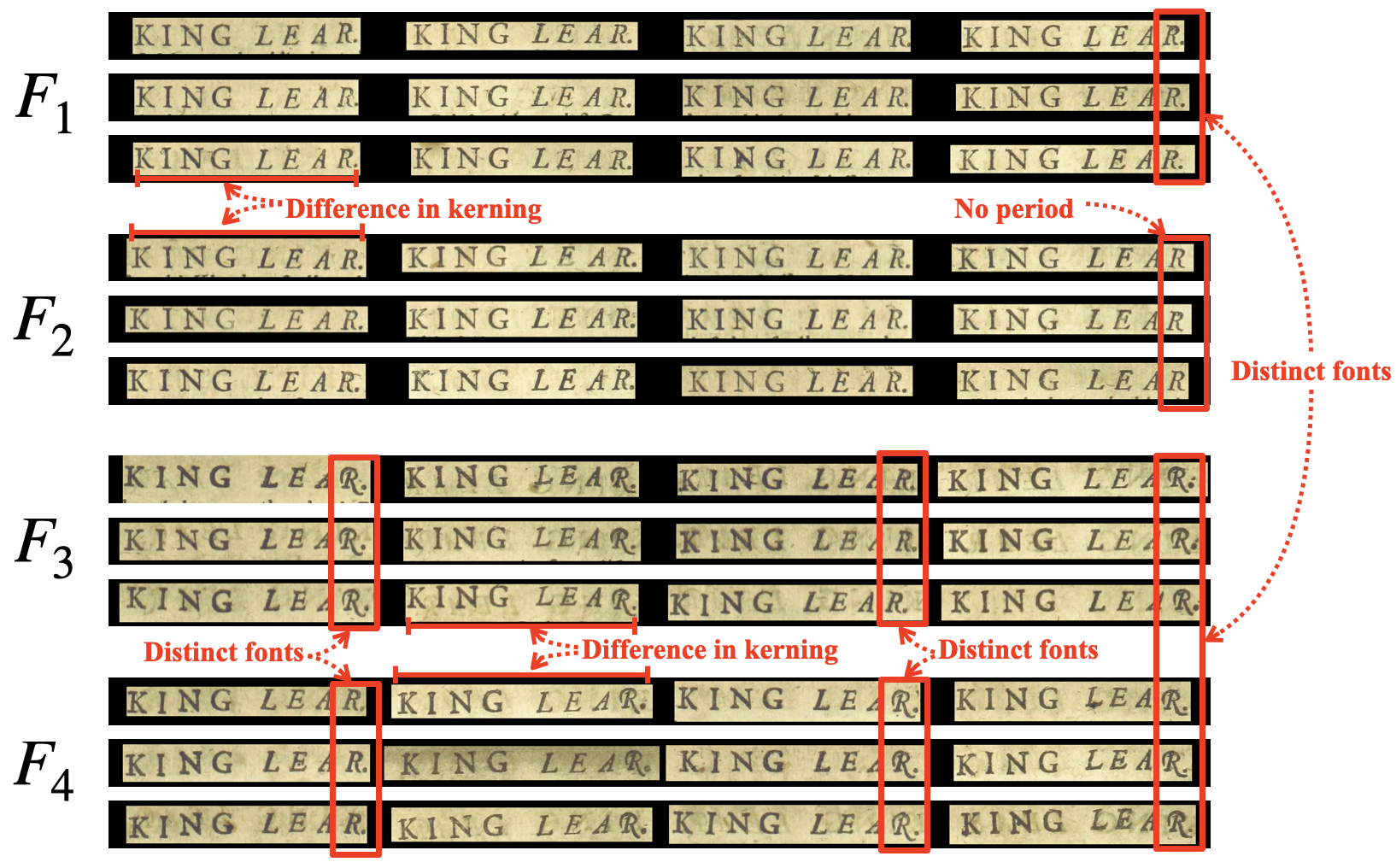}
    \caption{Annotated differences used to manually discover the latent clustering of skeleton formes for \textit{King Lear}, which was printed in the quarto format (i.e., 4 pages printed per side of a sheet). 
    Features that were used to discriminate between clusters during the annotation process are highlighted and described in red.  
    In this figure, each row represents a sheet side with each of the corresponding extracted headlines appearing in the same columns. 
    Our proposed sheet side clustering model perfectly partitions these skeleton forme clusters, as shown in Table~\ref{tab:main_results}.}
    \label{fig:king_lear_clustering}
    \vspace{-1.5em}
    
\end{figure}

In this section, we describe the creation of a suite of eight early modern book clustering problems with annotated, ground truth skeleton forme cluster assignments for automatically extracted headlines, along with additional format and gathering information.
First, we ask expert bibliographers to collect a set of books of different formats.
To make human annotation of headline clusters possible, we only considered books that also had high resolution scans available.
In Table~\ref{tab:dataset}, we show details of these eight books: 
\textit{The Mayor of Quinborough} (1661), \textit{Wisdom of God} (1691), \textit{Moral Discourse of Power} (1690), \textit{Parthenissa} (1676), 
\textit{Paradise Lost} (1667), \textit{The Institution of Gentleman} (1660) (henceforth \textit{Institution}), \textit{The History of King Lear} (1699), and \textit{Leviathan Ornaments} edition (1695, ESTC R13935).
Using signature marks at the bottom of pages and knowledge of the format of the books, we annotate the way the books were folded and gathered together.
Next, we extracted running titles from the headlines of each book automatically using the docTR scene text recognition model \citep{doctr2021}, and post-corrected $10\%$ of results to obtain clean running titles.
We then \textit{unfold} the sheet sides and create a montage of $M \times N$ sheet sides/pages per sheet side, such that every image in a column corresponds to a running title in the same position on their respective sheet side/skeleton forme used to print them.
Finally, expert bibliographers label true running title clusters for the eight books via prolonged examination of the high resolution images of the books to detect unique, minute variation in individual stamps used to print the titles.  
We show an example of this process for the book \textit{King Lear} in Figure~\ref{fig:king_lear_clustering}.
For headlines to be considered identical, the same sets of headlines need to move together for all the component pages, otherwise we consider the group of running titles as a new cluster.

\section{Results}\label{sec:results}

\begin{table}[t]
    \centering
        \caption{
    Clustering results (avg. of 5 runs) on our newly introduced suite of 8 ground truth clustering problems. 
    For each book, we report V-measure, 1-to-1 accuracy, and many-to-1 accuracy on recovering the underlying skeleton forme clustering partitions, given the true number of clusters. 
    Hyperparameters are chosen on the \textit{Paradise Lost} validation set, and the other books are treated as test books.
    }
    \label{tab:main_results}
    \begin{adjustbox}{width=\linewidth,center}
    \begin{tabular}{l | c c c | c c c | c c c | c c c}
    \toprule
         Method & & \textit{Mayor} & & & \textit{Discourse} & & & \textit{Parthenissa} &  & & \textit{Wisdom} &  \\
          & & (quarto) & & & (octavo) & & & (quarto) &  & & (octavo) &  \\
          & V & 1-1 & M-1  & V & 1-1 & M-1 & V & 1-1 & M-1 & V & 1-1 & M-1 \\
    \midrule
       Random Uniform & 37 & 39 & 53  & 1 & 54 & 54 & 52 & 31 & 37 & 38 & 36 & 40  \\
       Assign Majority &  0 & 39 & 39  & 0 & 50 & 50 & 0 & 13 & 13 & 0 & 30 & 30  \\
       Random Shuffle Labels & 44 & 47 & 61  & 0 & 52 & 52 & 52 & 32 & 39 & 39 & 37 & 41 \\
    \midrule
    \textit{ViT} Kernel - all pages & 13 & 34 & 44 & 0 & 51 & 51 & 18 & 18 & 22 & 6 & 26 & 31 \\
    \textit{ViT} Kernel - recto pages & 26 & 38 & 49 & 1 & 52 & 52 & 29 & 22 & 28 & 13 & 26 & 33 \\
    \textit{ViT} Kernel - sheet sides & 57 & 46 & 69 & 0 & 50 & 50 & 58 & 36 & 46 & 40 & 39 & 40  \\    
    \midrule
    \textit{Lev} Kernel - all pages & 23 & 36 & 51 & 0 & 51 & 51 & 45 & 29 & 39 & 15 & 27 & 36   \\
    \textit{Lev} Kernel - recto page & 48 & 44 & 72 & 8 & 67 & 67 & 56 & 37 & 47 & 36 & 35 & 49   \\
    \textit{Lev} Kernel - sheet side & \textbf{92} & \textbf{94} & \textbf{94} & \textbf{100} & \textbf{100} & \textbf{100} & \textbf{85} & \textbf{74} & \textbf{82} & \textbf{86} & \textbf{80} & \textbf{87}  \\
    \bottomrule
    \toprule
         Method & & \textit{Institution} & & & \textit{King Lear} &  & & \textit{Leviathan} & & & \textit{Paradise Lost} & \\
         & & (octavo) & & & (quarto) &  & & (folio) & & & (quarto) \\
    \midrule
       Random Uniform & 36 & 50 & 50 & 8 & 64 & 64 & 31 & 20 & 23 & 30 & 25 & 31 \\
       Assign Majority & 0 & 25 & 25 & 0 & 50 & 50 & 0 & 11 & 11 & 0 & 25 & 25  \\
       Random Shuffle Labels  & 34 & 47 & 47 & 18 & 72 & 72 & 30 & 19 & 23 & 29 & 25 & 31  \\
    \midrule
    \textit{ViT} Kernel - all pages & 1 & 52 & 52 & 13 & 37 & 38 & 15 & 15 & 18 & 8 & 21 & 27 \\
    \textit{ViT} Kernel - recto pages & 6 & 58 & 58 & 17 & 39 & 40 & 26 & 20 & 23 & 16 & 22 & 28   \\
    \textit{ViT} Kernel - sheet sides & 100 & 100 & 100 & 50 & 67 & 67 & 45 & 28 & 39 & 35 & 32 & 39  \\    
    \midrule
    \textit{Lev} Kernel - all pages  & 0 & 50 & 50 & 53 & 60 & 60 & 69 & 47 & 64 & 18 & 21 & 30 \\
    \textit{Lev} Kernel - recto page & 17 & 72 & 72 & 51 & 58 & 58 & 86 & 75 & 84 & 26 & 25 & 34  \\
    \textit{Lev} Kernel - sheet side  & \textbf{100} & \textbf{100} & \textbf{100} & \textbf{100} & \textbf{100} & \textbf{100} & \textbf{87} & \textbf{76} & \textbf{86} & \textbf{51} & \textbf{48} & \textbf{52} \\
    \bottomrule
    \end{tabular}
    \end{adjustbox}
    \vspace{-1em}
\end{table}

In Table~\ref{tab:main_results}, we present an evaluation of our approach by clustering the running titles from the eight book clustering problems into their ground truth skeleton forme assignments.
We use \textit{Paradise Lost} as our validation set to determine the appropriate hyperparameters.
In particular, we use quantile binning into $n=5$ bins, $p=4$-norm for our sheet combination strategy, and use $n=5$ neighbors and the same number of components as clusters for each book.

For each book, we report V-measure (V) \cite{rosenberg2007v}, 1-to-1, and many-to-1 accuracy \citep{christodoulopoulos2010two}.
The V-measure is a commonly used clustering metric defined as the harmonic mean between homogeneity and completeness, similar to the F1 score for classification, which is the harmonic mean between precision and recall.
For 1-to-1 accuracy, we map predictions to the ground truth clusters using the Hungarian algorithm.
The top rows of the table show simple baselines consisting of labeling sheet sides uniformly at random, assigning all labels to the cluster label with the greatest number of images, and randomly shuffling the true cluster assignments.
Below that, we present our main \textit{sheet side} approach, as described in our Approach section, which uses our custom-designed kernel function and the underlying format and folding structure of the book in order to build a similarity matrix.
Finally, we present a variation of this approach that does not use any structure information, instead relying on only the right-hand sides (i.e., \textit{recto}, as used by bibliographers) of each sheet side to perform the clustering, hence missing out on correlated sheet information.

First, we observe that both of the approaches for similarity outperform the random baselines indicating the effectiveness of our approach. Importantly, the additional bibliographic prior built into our sheet side model's kernel function significantly helps clustering accuracy.
We achieve perfect clusterings of three separate held-out books in different formats: \textit{Institution} and \textit{Discourse} in octavo, and \textit{King Lear} in quarto.
Also, performance is almost doubled compared to the recto model on \textit{Wisdom} (octavo) and  2/3 metrics for \textit{Paradise Lost} in quarto.
This result is intuitive because, in contrast with \textit{Leviathan}, which is printed in folio, the quarto and octavo format books have double and quadruple the amount of running titles that they can use to discriminate between skeleton forme clusterings.
In Figure~\ref{fig:king_lear_clustering}, we depict the clusterings for \textit{King Lear} to further prove the importance of this correlated skeleton forme information.
Whereas clear column-wise differences can be spotted for these running titles as highlighted, combining observations row-wise for each clustering decision can help break ties between these columns.

Finally, we observe that the neural kernel function consistently significantly under performs our domain-informed edit-distance based approach. Although, we fine-tuned a powerful pretrained vision transformer model which is theoretically more expressive than a simple feature-based model, we observed that the ability to directly incorporate domain knowledge about spacing and shape differences via our edit-based approach is crucial compared to only being able to incorporate the domain knowledge into our neural method indirectly via synthetic data augmentation.
\begin{figure}[ht!]
    \centering
    \includegraphics[width=\textwidth]{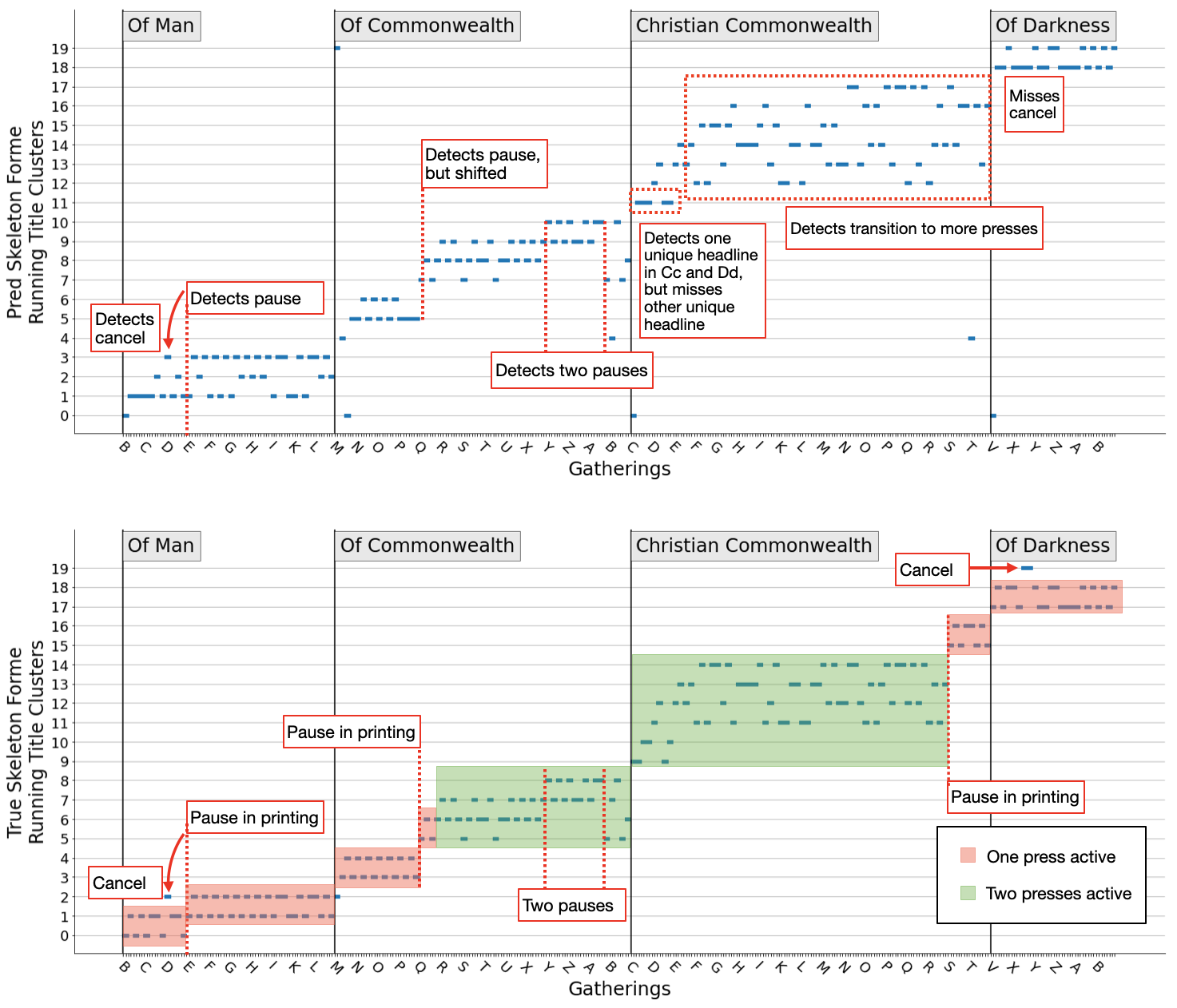}    
    \caption{We show ground truth (bottom) and predicted (top) skeleton forme cluster assignments as they appear across the four sections (\textsc{Of Man}, \textsc{Of Commonwealth}, \textsc{Christian Commonwealth}, \textsc{Of Darkness}) of \textit{Leviathan} from the start of the book on the left to the finish on the right. On the x-axis, we label the gathering IDs as they appear (and repeat themselves) throughout the book. Each blue dot represents a page, but they occur in contiguous pairs of two due to our unfolding of its folio format (two pages per sheet side). For the ground truth, we summarize what the different skeleton forme usage reveals about the making of the book, as initially revealed in \citep{warrenetal2021}. Finally, we annotate on the top plot where our predicted clusters succeed and fail against the ground truth findings. See Qualitative Findings section for more details.}
    \label{fig:staircase}
    \vspace{-2em}
\end{figure}

\subsection{Qualitative Bibliographic Findings}\label{sec:staircase}
On top of numerical measures of accuracy, in Figure~\ref{fig:staircase} we present a detailed error analysis that describes our method's successes and failures in predicting important bibliographic details on the held-out \textit{Leviathan} book.
We show ground truth (bottom) and predicted (top) skeleton forme cluster assignments as they appear across the four sections (\textsc{Of Man}, \textsc{Of Commonwealth}, \textsc{Christian Commonwealth}, \textsc{Of Darkness}) of \textit{Leviathan} from the start of the book on the left to the finish on the right. 
On the y-axis we denote the 20 different skeleton forme cluster IDs, while on the x-axis, we label the letter IDs of the gatherings of sheets as they appear (and repeat themselves) throughout the book. 
We note that each individual blue dot represents a page, but the pages occur in contiguous pairs due to our unfolding of its folio format (two pages per sheet side).
For the ground truth, we summarize what the different skeleton forme usage reveals about the making of the book, as initially revealed in \citep{warrenetal2021}.
For instance, we highlight (in green) regions where two printing presses were probably used in parallel to produce the book, which can be inferred from the multiple active skeleton formes that alternate across those sections.
We also show pauses, in which a stop press was probably declared due to the skeleton formes potentially being used for other purposes.
Finally, we indicate the presence of two cancels in the first and last sections of the book, indicative of revised errors in printing.

When we compare our sheet side model predictions (Figure~\ref{fig:staircase}, top) to the ground truth below it, we can see that our clustering correctly detects many of the bibliographically relevant features that are found in the ground truth clusters.  
For instance, it correctly detects the cancel in \textsc{Of Man}, as well as the headline interruption in that section.  
It also correctly detects all three headline interruptions in \textsc{Of Commonwealth}, though one is slightly off in location.  
It detects one of the two unique headlines in the first two gatherings of \textsc{Christian Commonwealth}, as well as the transition to more presses in that section.
However, it assigns too many clusters to the later half of \textsc{Christian Commonwealth}, and misses the cancel anomaly in \textsc{Of Darkness}.
We hypothesize that such cancels may be difficult to detect with the $k$-means algorithm we use, as it is not incentivized to explain a single, anomalous point with a cluster.

\section{Related Work: Analytical \& Computational}
\paragraph{\textbf{Analytical Bibliography:}} 
The manual identification and tracking of skeleton formes for analyzing early modern presswork was first established by Alfred Pollard in 1909 \citep{pollard1909shakespeare}, followed by more systematic analyses \cite{pollard1909shakespeare,willoughby1928note,bowers1938notes}.
Since Fredson Bowers popularized headline analysis in the mid-twentieth century, scholars have used the technique to uncover details of the printing of Shakespeare's \textit{Hamlet} and \textit{Sonnets} \cite{eganPressVariantsQ22015, mcleodTechniqueHeadlineAnalysis1979}.
Rees' discovery \citep{rees2009publishing} that at least one book printed by seventeenth-century printer John Bill used eight separate skeleton formes refuted the long-standing theory that early modern print shops used, at most, one or two skeleton formes concurrently  \cite{rees2009publishing}.
Recent manual analysis of \textit{Leviathan} also shows proliferating skeleton formes \cite{warrenetal2021}.
Tracking such a large amount of skeleton formes across a book's pages takes considerable time and effort, and techniques such as photocopied transparency overlays have been designed to compare headlines more rapidly \cite{mcleodTechniqueHeadlineAnalysis1979}.  
However, even the most successful of such techniques remain manually intensive due to the immense human effort required.

\paragraph{\textbf{Computational Methods:}} Recently, some bibliographers have moved from analysis of physical copies of books to their high-resolution digital facsimiles and have started to leverage computational methods in their investigations \citep{stahmer_digital_2016, piperPageImageVisual2020, abuelwafaDetectingFootnotes322018a, zhalehpourVisualInformationRetrieval2019, mhiriFootnotebasedDocumentImage2017}.
Machine learning approaches for the digital humanities play an integral role in these advancements.
For instance, early modern optical character recognition (OCR) allows for automatic extraction of text content \citep{Vogler2023Contrastive}, while probabilistic models for typographical analysis enable study of their fonts \citep{goyal2020probabilistic}.
Our work is most related to automatic compositor attribution methods for Shakespeare's First Folio \citep{ryskina2017automatic}, which clustered book pages by the individuals who set the type using orthographic visual features.
Our contribution is in many ways complementary to this approach as it clusters portions of the book using spacing-aware features, which were one of the most discriminative orthographic features from \citep{ryskina2017automatic}.
In contrast to identifying individuals who set the type using all of the recognized body text, however, we identify the underlying skeleton forme templates used in the printing of the book.
To accomplish this, we only use the running titles in low-resolution images of the headlines of a book using features that do not rely on OCR output.

\vspace{-1em}
\section{Conclusion}
We presented a new approach to cluster small variations in running titles of early modern printed books in order to provide evidence of how they were printed.
We attacked the problem from two different angles: building domain knowledge into a similarity kernel directly and learning a neural cross-encoder similarity kernel indirectly via data synthesis and augmentation.
We found that the former performs much better than the latter, which could be due to the subtle distinctions in the headlines and noise present in early modern printed books that can be difficult to learn from fake, rendered data.
We also leveraged the underlying printing format and folding structure of the book in our method by defining reduction operations that combine correlated parts of the skeleton formes.
Quantitative results on a newly annotated set of eight early modern books demonstrates the overall promise of our method, while a qualitative comparison to an extensive manual investigation of Hobbes' \textit{Leviathan} shows the bibliographic evidence it can automatically uncover.
Because our approach mainly considers macroscopic kerning variations and font deviations, application of our methods hold promise for bibliographical analysis of abundantly available low-resolution scan images of books, such as the ones available in the Early English Books Online (EEBO) collection.

\vspace{-0.8em}
\section{Limitations}
Analyzing headlines relies on books being printed with a small group of the same, repeated skeleton formes, which may not exist for books with frequently reset running titles, such as the influential King James Bible (1611).
In this case, existing automatic compositor attribution techniques \citep{ryskina2017automatic}, which use features derived from body text OCR to discover the number of compositors, could be used in tandem with our approach for robustness. 
Additionally, our method relies on knowledge of the ground truth number of clusters for k-means clustering, but this can be remedied using non-parametric style approaches for clustering spectral embeddings, such as Dirichlet Process GMMs with Chinese Restaurant Process priors.

\section*{Acknowledgments}
This work was supported in part by the Mellon Foundation (Award No. 307964-00001), along with grants from the National Science Foundation (Print and Probability: A Statistical Approach to Analysis of Clandestine Publication; NSF 1816311), XSEDE (HUM150002), and the National Endowment for the Humanities (Freedom and the Press before Freedom of the Press: Tools, Data, and Methods for Researching Secret Printing; HAA-284882-22).

\newpage
\appendix
\section{Supplementary Material}
\begin{figure}[h!]
    \centering
    \includegraphics[width=0.85\textwidth]{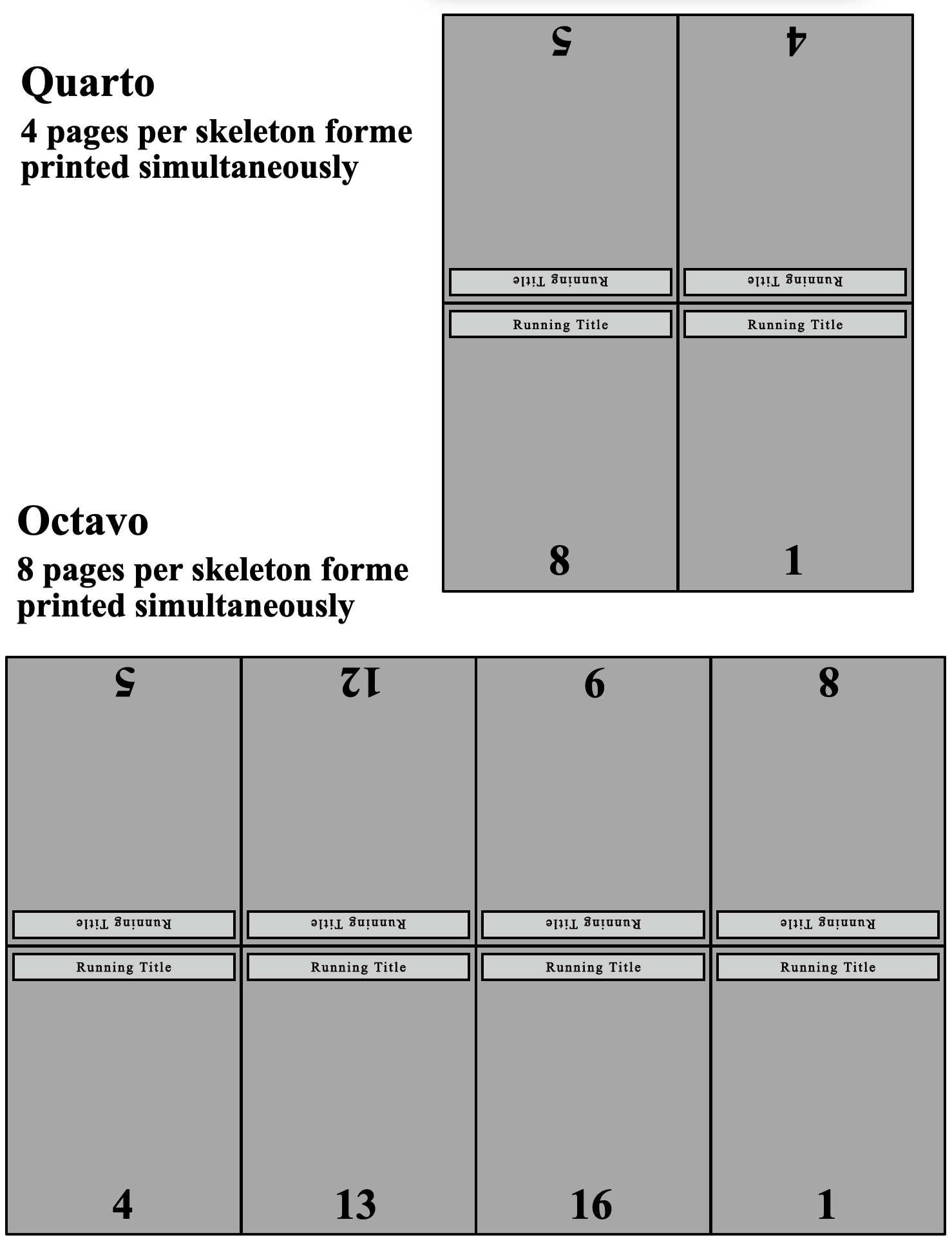}
    \caption{
    While a \textit{folio} (not pictured here, see Fig.~\ref{fig:overview}) was printed using only two pages per side of a sheet, \textit{quarto} format involved printing four pages per sheet side, and \textit{octavo} format eight pages per sheet side, etc. We show the skeleton formes for these formats, and label the page numbers as well as where the running titles would be (although they would be mirrored on an actual skeleton forme, as shown in Fig.~\ref{fig:overview}.
    Each format, diminishing by size, required an increasingly complex pattern of \textit{imposition}, or the arrangement in which sheets were set on the press to yield a sequentially-ordered gathering of pages after folding. Figure inspired by \citep{werner2019studying}.
    }
    \label{fig:quarto_octavo}
\end{figure}

\bibliographystyle{splncs04}

\end{document}